\documentclass[doublespacing]{elsart}
\usepackage{natbib}
\usepackage{graphicx}
\begin{document}
\begin{frontmatter}
\title{Electron density diagnostic potential of Ar XIV soft X-ray emission lines}
\author{G. Y. Liang\corauthref{Liang}}~$^{\ast}$,
\corauth[Liang]{Corresponding author: Gui-Yun Liang}
\ead{gyliang@bao.ac.cn}
\author[Liang]{G. Zhao},
\author[Liang,Zeng]{J. L. Zeng},
\author[Liang]{J. R. Shi}
\address[Liang]{National Astronomical Observatories, Chinese
Academy of Sciences, Beijing
 100012, P. R. China}
\address[Zeng]{Department of Applied Physics, National University
of Defense Technology,
        Changsha 410073, P. R. China}

\begin{abstract}
Theoretical electron density-sensitive line ratios $R_1 - R_6$ of
Ar XIV soft X-ray emission lines are presented. We found that
these line ratios are sensitive to electron density $n_e$, and the
ratio $R_1$ is insensitive to electron temperature $T_e$. Recent
work has shown that accurate atomic data, such as electron impact
excitation rates, is very important for reliable determination of
the electron density of laboratory and astrophysical plasmas.
Present work indicates that the maximum discrepancy of line ratios
introduced from different atomic data calculated with distorted
wave and R-matrix approximations, is up to 18\% in the range of
$n_e=10^{9-13}$cm$^{-3}$. By comparison of these line ratios with
experiment results carried out in electron beam ion trap
(EBIT-II), electron density of the laboratory plasma is diagnosed,
and a consistent result is obtained from $R_1$, $R_2$ and $R_3$.
Our result is in agreement with that diagnosed by Chen \etal using
triplet of  N VI. A relative higher diagnosed electron density
from $R_2$ is due to its weak sensitivity to electron temperature.
A better consistency at lower $T_e$ indicates that temperature of
the laboratory plasma is lower than log$T_e$(K)=6.5. Comparison
between the measured and theoretical ratios reveals that
32.014~\AA\, line is weakly blended by lines from other Ar ions,
while 30.344~\AA\, line is strongly contaminated.
\end{abstract}
\end{frontmatter} \maketitle

\section{Introduction}
Spectroscopy has exhibited its great importance for diagnostic of
laboratory and astrophysical plasmas. Emission lines arising from
$n=2\to n=2$ transitions of B-, C- and N-like ions have been
frequently observed in solar extreme-ultraviolet spectra~[1]. They
have been extensively used to determine electron temperature
and/or density of stellar coronae through line intensity ratios.
Keenan \etal [2-4] presented electron density-sensitive ratios for
B-like ions including Si X, Ar XIV and Ca XVI etc. For C-like ions
including S XI, Ca XV, Fe XXI and Ar XIII, the electron
density-sensitive ratios were also given by Keenan \etal [5-8].
For N-like ions, to our best knowledge, the electron
density-sensitive ratios were presented only for S X~[9]. Using
the ratios from these ions' lines, accurate electron densities
have been obtained, which are consistent with those from iron ions
with same temperatures of maximum fractional abundance in
ionization equilibrium~[10]. Another benefit using these line
ratios is that the uncertainty due to ionization equilibrium and
element abundance has been eliminated.

With the launch of a new generation of X-ray satellites, including
{\it Chandra} and {\it XMM}, a large amount of high quality
spectra with high spectral resolution and high effective area have
been obtained for nearly all classes of astrophysical X-ray
sources~[11-17]. This will allow us to estimate physical
parameters such as the electron temperature and density of the
stellar X-ray corona. The temperature has long been available
before the launch of the two satellites, while the density
determination of X-ray emitting layer becomes possible until the
launch of satellites with the high-spectral resolution. Spatial
information of stars could be assessed indirectly from
relationship between electron density and emission measure
($EM=n_e^2V$). For reliable determination of the electron density
and temperature, accurate atomic data is very important,
especially electron impact excitation rates and oscillator
strengths~[8].

Currently, the electron density is mainly determined using triplet
line ratios (resonance, inter-combination and forbidden lines) of
He-like ions for line formation regions. Ness \etal [18] and Testa
\etal [19] present an extensive investigation using those He-like
ions for stars with different activity levels, and disclose that
the electron density of stellar coronae does not exceed
$5\times10^{12}$cm$^{-3}$. One reason of the extensive application
of He-like ions is that a larger amount of reliable atomic data is
available, because the atomic data calculation of K-shell ions is
relatively simple. And some laboratory experiments conducted in
Tokamak, electron beam ion traps (EBIT) and intense laser-matter
interaction have been carried to benchmark the theoretical
calculation for K-shell ions~[20,21]. Many L-shell emission lines
of iron ions have been detected for star coronae. The temperature
structure of the X-ray emitting layers have been obtained using
these iron lines. However, further investigation is still
necessary to depict out the structure of the stellar coronae and
to explain heating mechanism.

In X-ray spectra of stars with various activity, the L-shell
emission lines of non-iron elements including Si, S, Ar and Ca
have also been detected. But recent studies revealed that there
are large uncertainties for the atomic data of these ions.
Fortunately, a large project---Emission Line Project supported by
the Chandra X-ray Observatory, has been opened to complete a
comprehensive catalog of astrophysically relevant emission lines
in support of new-generation X-ray observatories. Lepson
\etal~[22] recently made an experiment for highly charged Ar ions
in EBIT-II to benchmark theoretical calculation. The intensity
ratios of the L-shell emission lines of these non-iron ions might
be used for direct density estimations of stellar coronae and
fusion plasma. Therefore the exploration and calibration of the
line intensity ratios of the L-shell emission lines of the
non-iron elements, such as Si, S, Ar and Ca, are necessary in the
next few years.

Argon is an abundant element, and has been detected in solar and
solar-like stars, such as Capella~[13], Algol~[23] and
AU~Mic~[24]. For example, the emission lines arising from $n=3-2$
transitions of Ar XVI have been identified by Audard \etal~[13],
and $3d - 2p$ transition lines located near 27.043~\AA\, of Ar XV
have also been identified in our previous work~[25]. In ionization
equilibrium condition~[10], the temperature of maximum Ar XIV
fractional abundance has a same value with that of He-like Ne.
However the triplets of He-like Ne are strongly blended by
forest-like features of highly charged Fe ions~[26]. So the
features of Ar XIV may give the density information in Ne IX ion
formation region. At present, APEC database is the most complete
and has been extensively used in X-ray spectral analyses for
stars. In fact APEC extends Chianti database by including some
experimental values of highly charged iron ions. In Chianti
database, emission lines of Ar XIV are also included. However all
the collision rates are results withs distorted wave (DW)
approximation and using less configurations. Therefore the
exploration of the density-sensitive line ratios of highly charged
Ar XIV using more accurate atomic data will be meaningful.

In this paper we present calculations of electron
density-sensitive line intensity ratios of Ar XIV in X-ray line
formation region. Present calculation of atomic structure includes
more configurations than previous work, and some excitation rates
are replaced by data from R-matrix method. By comparing these
ratios with experimental ones conducted by Lepson \etal~[22] in
the Lawrence Livermore electron beam ion traps EBIT-II, the
electron density of the laboratory plasma is determined.

\section{Theoretical line ratios}
The model ion for Ar XIV consists of 152 fine-structure energy
levels belonging to the configurations $2s^22p$, $2s2p^2$, $2p^3$,
$2s^23l$, $2s2p3l$, $2p^23l (l=s, p, d)$, $2s^2nl$~($n=4,5,6$
and$l=n-1$).

Electron impact excitation rates are obtained by performing
numerical integration of collision strengths over a Maxwellian
distribution. The collision strengths among the lowest 152 energy
levels, along with the spontaneous radiative decay rates and
energy levels were calculated using the Flexible Atomic Code (FAC)
developed by Gu~[27,28]. The atomic structure calculation in FAC
is based on the fully relativistic configuration interaction
approximation, while the electron impact collision strengths are
calculated in the relativistic DW approximation. These data are
available electronically from the authors (gyliang@bao.ac.cn) on
request. For the electron impact excitation rates among the lowest
15 levels, we adopted recent results of Keenan \etal~[8], in which
a more accurate method--- R-matrix (Burke and Roob~[29]) is
adopted. As noted by, for example, Seaton~[30], excitation by
protons will be important for the lowest levels. This will affect
the X-ray emission lines indirectly. In the present analysis we
have employed the rates of Foster, Keenan $\&$ Reid~[31], which
were calculated using a close-coupled impact-parameter method.

Using the atomic data discussed above in conjunction with rate
equations, relative Ar XIV level populations were derived for a
range of electron temperatures and densities. The rate equations
are given as~[25,32,33]
\begin{eqnarray}
N_j[\sum_{i<j}A_{ji}+n_e(\sum_{i<j}C_{ji}^d+\sum_{i>j}C_{ji}^e)] &
= &
\sum_{i>j}N_iA_{ij}+n_e(\sum_{i<j}N_iC_{ij}^e+\sum_{i>j}N_iC_{ij}^d)
\nonumber,
\end{eqnarray}
where the superscript $e$ and $d$ refer to electron excitation and
deexcitation, $N_j$ is the number density of level $j$, and $n_e$
is the electron density. In the calculations, following
assumptions were made: (i) photon-excitation and de-excitation
rates are negligible in comparison with the corresponding
collision rates; (ii) ionization to and recombination from other
ionic levels are slow compared with bound-bound rates; (iii) the
plasma is optically thin. Generally, the line intensity for a
given transition line $\lambda_{ji}$ is given by
\begin{eqnarray}
I_{ji} & = & N_jA_{ji}.\nonumber
\end{eqnarray}

In order to correctly understand the astrophysical spectra, Lepson
\etal~[22] carried out an experiment in Lawrence Livermore
electron beam ion traps EBIT-II for Ar at three different energies
of 600, 650 and 1000~eV. This is part of a large project to
complete a comprehensive catalog of astrophysically relevant
emission lines in support of new-generation X-ray observatories.
In this experiment, several emission lines of Ar XIV were
detected, as shown in Table 1.
\begin{table*}
\caption[1]{Experimental wavelength~[22] and line identification
of Ar XIV. The theoretical intensity ratios are calculated at a
temperature (log$T_e$(K)=6.5) and density
(5.6$\times$10$^{10}$~cm$^{-3}$). The intensities are normalized
to that of 27.469~\AA\, line.}
    \vspace{0.1cm}
\[
\hspace{-0.6cm} \begin{array}{lllllll}
            \hline
\multicolumn{3}{c}{\rm Wavlength~(\AA)} & \multicolumn{3}{c}{\rm Relative~intensity} &  \\
 {\rm Exp.} & {\rm HULLAC} & {\rm FAC} & {\rm Measured} & {\rm HULLAC} & {\rm FAC} & {\rm
Transition} \\
            \hline
{\rm 27.469} & {\rm 27.517} & {\rm 27.470} & {\rm 1.00} & {\rm 1.00} & {\rm 1.00} & 2s^23d_{3/2}-2s^22p_{1/2} \\
{\rm 27.631} & {\rm 27.674} & {\rm 27.629} & {\rm 0.95} & {\rm 1.25} & {\rm 0.76} & 2s^23d_{5/2}-2s^22p_{3/2} \\
...          & {\rm 27.688} & {\rm 27.642}^b & ...      & {\rm 0.19} & {\rm 0.20} & 2s^23d_{3/2}-2s^22p_{3/2} \\
{\rm 28.223} & {\rm 28.332} & {\rm 28.329} & {\rm 0.20} & {\rm 0.25} & {\rm 0.25} & ((2s2p_{1/2})_13d_{5/2})_{5/2}-((2s2p_{1/2})_12p_{3/2})_{3/2} \\
{\rm 30.215} & {\rm 30.342} & {\rm 30.236} & {\rm 0.20} & {\rm 0.25} & {\rm 0.20} & ((2s2p_{3/2})_23s)_{3/2}-((2s2p_{1/2})_12p_{3/2})_{5/2} \\
{\rm 30.344} & {\rm 30.476} & {\rm 30.367} & {\rm 0.15} & {\rm 0.25} & {\rm 0.41} & ((2s2p_{1/2})_13s)_{1/2}-((2s2p_{1/2})_12p_{3/2})_{3/2} \\
-            & -            & {\rm 31.360} & -          & - & - & ((2s2p_{1/2})_13s)_{1/2} - 2s(2p^2)_0 \\
{\rm 32.014} & {\rm 32.161} & {\rm 32.071} & {\rm 0.30} & {\rm 0.19} & {\rm 0.21} & 2s^23p_{3/2} - ((2s2p_{1/2})_12p_{3/2})_{5/2} \\
...          & {\rm 32.215} & {\rm 32.125}^b & ...        & {\rm 0.13} & {\rm 0.24} & 2s^23p_{1/2} - ((2s2p_{1/2})_12p_{3/2})_{3/2} \\
           \hline
         \end{array}
 \]
\flushleft{$^b$ denotes a blended line}
\end{table*}
Theoretical predictions of line intensities, including HULLAC [22]
and FAC calculations by us, are also presented.

By resolving rate equations, we found that the following emission
line intensity ratios of Ar XIV ion, are sensitive to electron
density:
\begin{eqnarray}
R_1 &= I(27.631~{\rm \AA})/I(27.469~{\rm \AA}), \nonumber \\
R_2 &= I(27.631~{\rm \AA})/I(28.223~{\rm \AA}), \nonumber \\
R_3 &= I(27.631~{\rm \AA})/I(30.215~{\rm \AA}), \nonumber \\
R_4 &= I(27.631~{\rm \AA})/I(30.344~{\rm \AA}), \nonumber \\
R_5 &= I(27.631~{\rm \AA})/I(31.360~{\rm \AA}), \nonumber \\
R_6 &=  I(27.631~{\rm \AA})/I(32.014~{\rm \AA}). \nonumber
\end{eqnarray}
Therefore they may be useful for the electron density diagnostics
for all kinds of hot plasma with sufficient signal-to-noise ratio.
Figures 1---4 show the theoretical line intensity ratios as a
function of the electron density at temperature (log$T_e$(K)=6.5)
of maximum Ar XIV fractional abundance in ionization
equilibrium~[10]. Predictions of ratios $R_3$ and $R_5$ are much
larger than one, so the determination of electron density will be
very difficult due to extremely faint lines involved.
\begin{figure*}
\centering
\includegraphics[angle=-90,width=8cm]{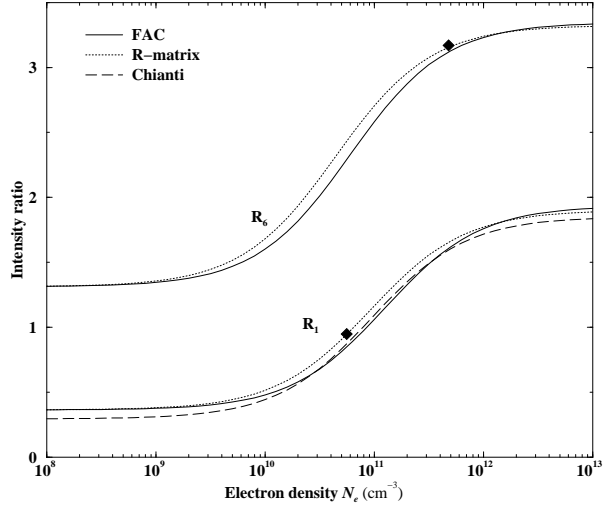}
\caption[short title]{The theoretical Ar XIV emission line ratios
$R_1$ and $R_6$, plotted as a function of electron density ($n_e$
in cm$^{-3}$) at the temperature of maximum Ar XIV fractional
abundance in ionization equilibrium, log$T_e$(K)=6.5. Filled
diamonds and Dashed curve are experimental values measured in
EBIT-II by Lepson \etal~[22], and Chianti prediction,
respectively.}
\end{figure*}
\begin{figure*}
\centering
\includegraphics[angle=-90,width=8cm]{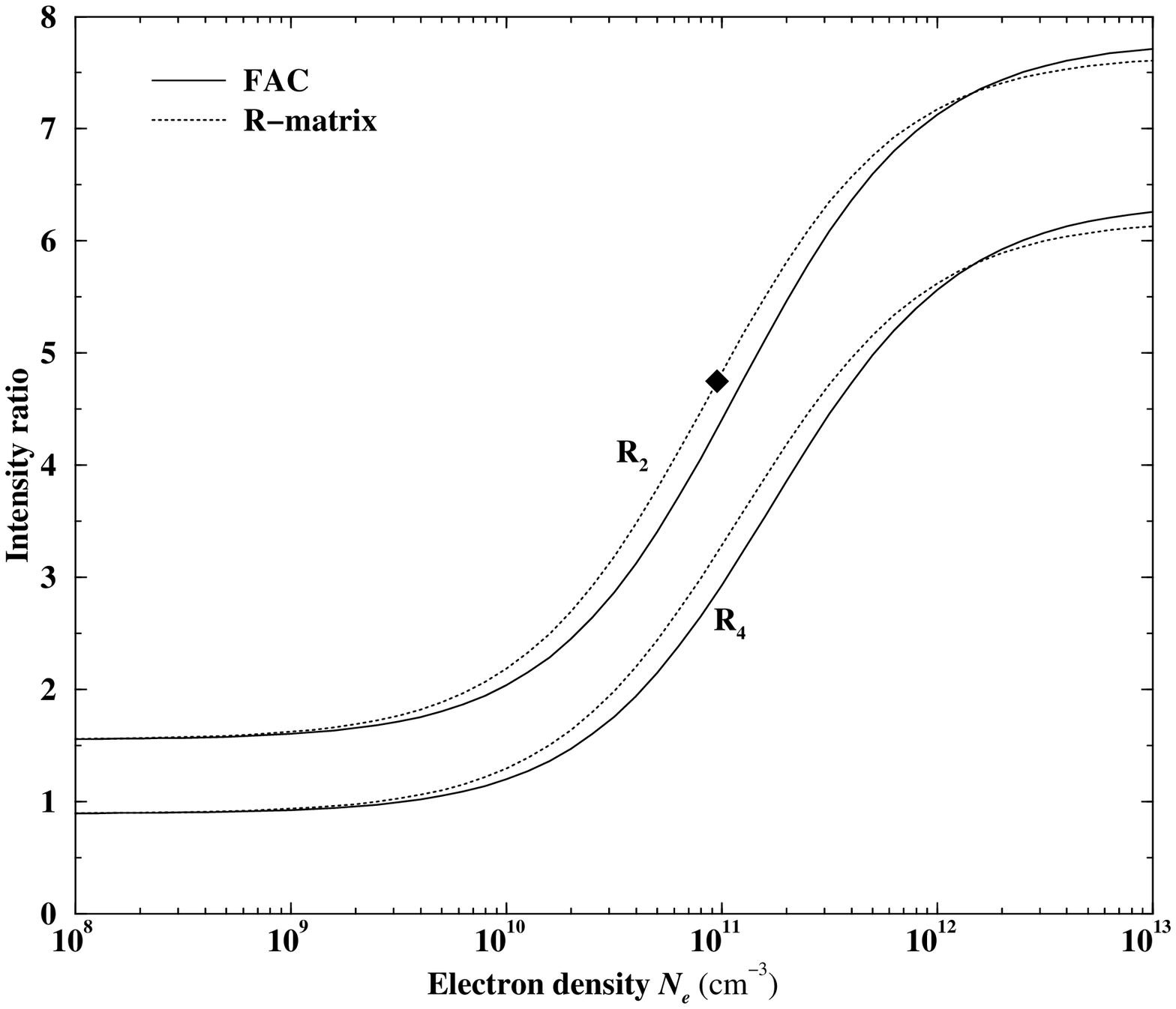}
\caption[short title]{The theoretical Ar XIV emission line ratios
$R_2$ and $R_4$, plotted as a function of electron density ($n_e$
in cm$^{-3}$) at the temperature of maximum Ar XIV fractional
abundance in ionization equilibrium, log$T_e$(K)=6.5. Filled
diamond are experimental values measured in EBIT-II by Lepson
\etal~[22].}
\end{figure*}
\begin{figure*}
\centering
\includegraphics[angle=-90,width=8cm]{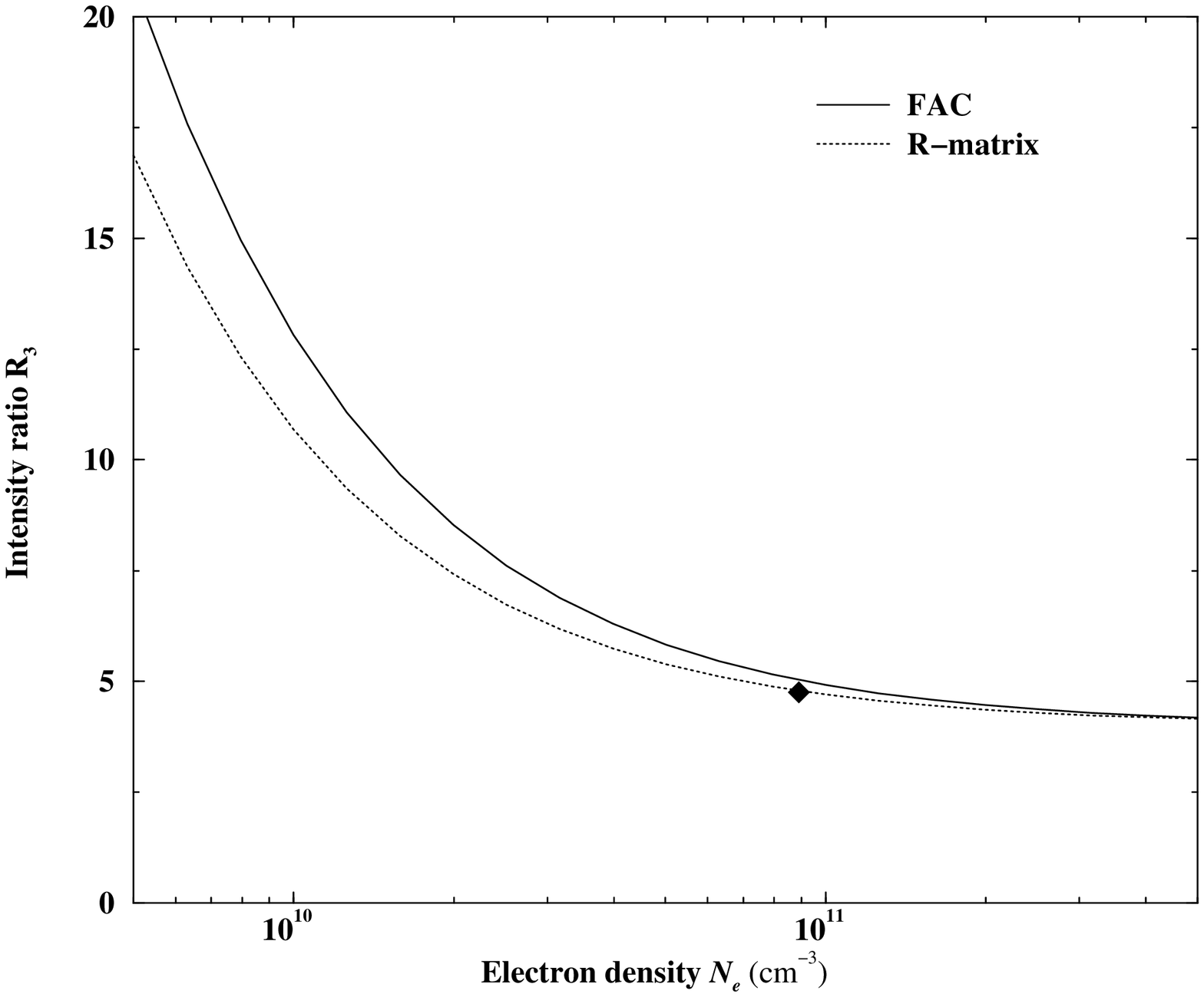}
\caption[short title]{The theoretical Ar XIV emission line ratios
$R_3$, plotted as a function of electron density ($n_e$ in
cm$^{-3}$) at the temperature of maximum Ar XIV fractional
abundance in ionization equilibrium, log$T_e$(K)=6.5. Filled
diamond are experimental values measured in EBIT-II by Lepson
\etal~[22].}
\end{figure*}
\begin{figure*}
\centering
\includegraphics[angle=-90,width=8cm,clip]{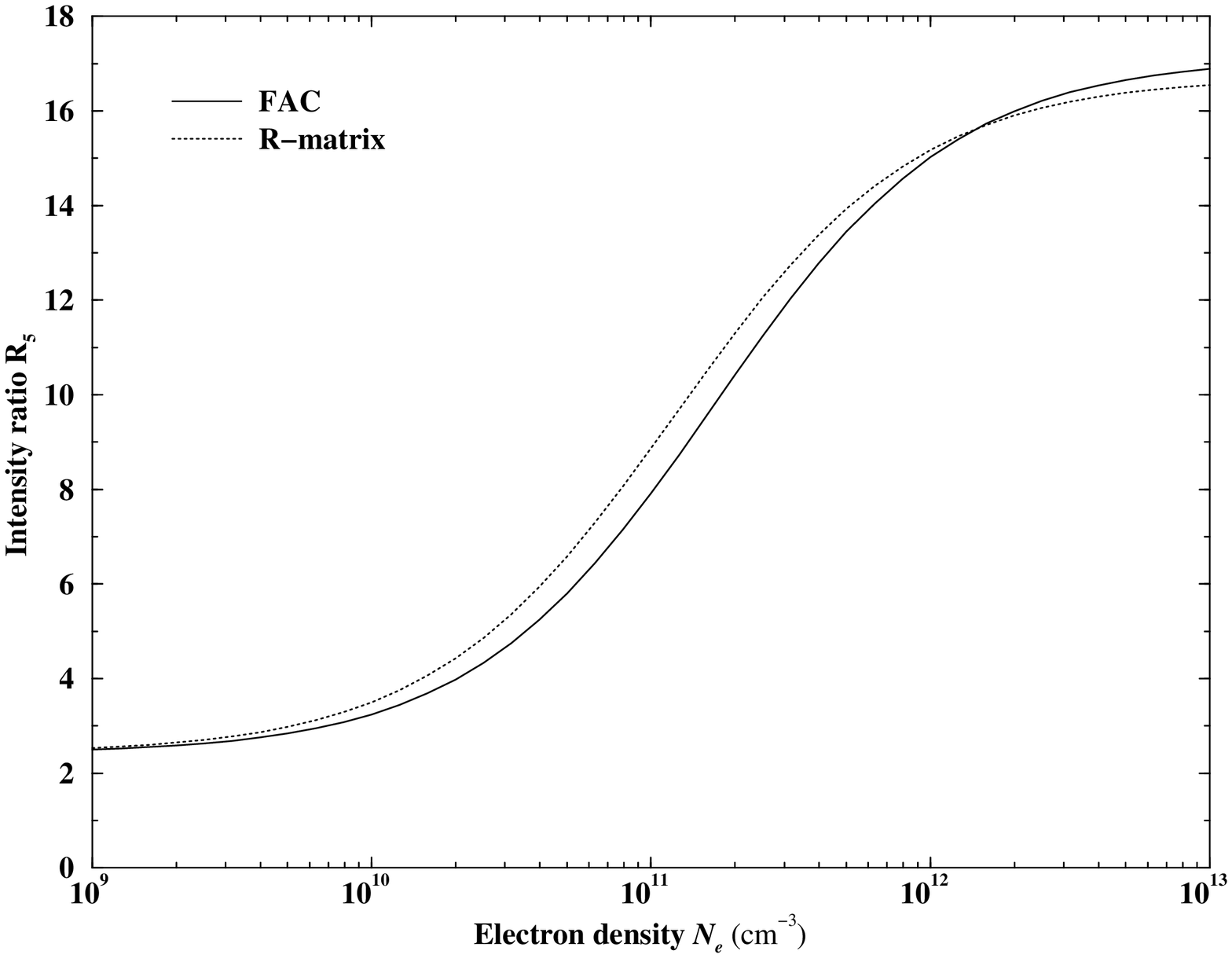}
\caption[short title]{The theoretical Ar XIV emission line ratios
$R_5$, plotted as a function of electron density ($n_e$ in
cm$^{-3}$) at the temperature of maximum Ar XIV fractional
abundance in ionization equilibrium, log$T_e$(K)=6.5.}
\end{figure*}
In these figures, the solid lines refer to results using the
excitation rates obtained from DW approximation, while the dotted
lines refer to results using the excitation rates for the lowest
15 energy levels obtained from R-matrix approximation~[8]. The
discrepancies between DW and R-matrix calculation reach up to 15\%
at $n_e$ = $10^{12}$ cm$^{-3}$, which indicates that the resonance
excitation effect is important in the calculation of the electron
impact excitation rates. Chianti prediction exhibits a similar
result with the FAC prediction in density-sensitive region as
shown in Fig. 1. For clarity, Chianti prediction is shown only for
ratio $R_1$. APEC shows similar result with Chianti prediction due
to same data adopted for Ar XIV. An inspection of the figures also
tells us that the ratios are sensitive to the electron density.
For example $R_1$ and $R_2$, vary by factors of 4.4 and 4.3
respectively, at $n_e$ = $10^9$ and $10^{12}$ cm$^{-3}$. Moreover
the ratio $R_1$ is insensitive to electron temperature. For
example, a change in $T_e$ of 0.2~dex (i.e. 58\%) leads to a less
than 1\% variation at $n_e=10^9$ cm$^{-3}$, up to 3\% at
$n_e=10^{11}$ cm$^{-3}$, as show in Fig. 5.
\begin{figure*}
\centering
\includegraphics[angle=-90,width=8cm,clip]{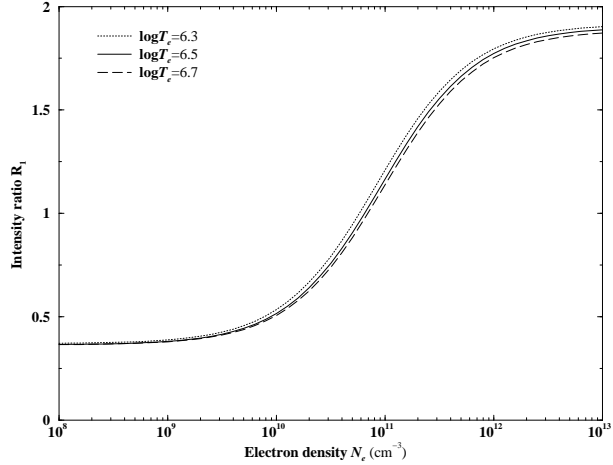}
\caption[short title]{The theoretical Ar XIV emission line ratio
$R_1$, plotted as a function of electron density ($n_e$ in
cm$^{-3}$) at temperature log$T_e$(K)=6.3 (dotted line), 6.5
(solid line) and 6.7 (dashed line)}
\end{figure*}
The sensitivity of the line intensity ratio $R_1$ to electron
density, combined with its insensitivity to $T_e$, makes it very
useful for the density diagnostics for laboratory and
astrophysical plasmas.

\section{Results and discussion}
Applying present line intensity ratios, we diagnose the electron
density of Ar laboratory plasma conducted by Lepson \etal~[22] in
EBIT-II. In the experiment, energy of electron beam is fixed at
1.0~KeV. However, the energy will become lower than 1.0~KeV when
the electrons interact with cold Ar gas in trap region. The
temperature of produced Ar plasma is not given by Lepson \etal
[22]. For detailed information of the experiment, you can refer
the paper written by Lepson \etal~[22]. Recently, Chen \etal~[34]
measured the electron density ($6.0\times10^{10}$cm$^{-3}$) of the
laboratory plasma using triplet line intensity ratio of He-like N
ion.

In Table 1, we present the measured intensities relative to that
of the line at 27.469~\AA\,, along with the theoretical results
including the HULLAC~[22] and FAC calculations by us. In the
theoretical calculation, the electron temperature adopted here is
log$T_e$(K)=6.5 which corresponds to maximum Ar XIV fractional
abundance in ionization equilibrium~[10], while the electron
density is $n_e=5.6\times10^{10}$cm$^{-3}$. In the ionization
equilibrium condition, the fraction of Ar XIV as a function of
temperature has a steep peak at log$T_e$(K)=6.5. Since emissivity
almost comes from plasma with this temperature, we assume the
produced plasma is isothermal plasma with temperature
log$T_e$(K)=6.5. Table 1 shows that our results are in agreement
with experimental measurements and are better than the HULLAC
calculation in wavelength. For the relative intensities, our
calculation also agree with experimental measurement except for
the line at 30.344~\AA\,.

Above section has shown that the ratio $R_1$ is sensitive to the
electron density $n_e$, while it is insensitive to the electron
temperature $T_e$. Obviously this ratio is a good diagnostic tool
for the electron density $n_e$. Moreover, the laboratory
measurement and theoretical calculation indicate that emission
lines relevant to the ratio $R_1$ are the strongest lines of Ar
XIV and not contaminated by other Ar lines. So we can use $R_1$ to
diagnose the electron density of the laboratory plasma. By
comparison between the measured value and theoretical calculation
as shown by the filled diamond in Fig. 1, the electron density
$n_e=5.6\times10^{10}$cm$^{-3}$ is derived. This result is
consistent with value $6.0\times10^{10}$cm$^{-3}$ recently
diagnosed by Chen \etal~[34] using triplet line intensity ratio of
He-like N ion.

Table 2 shows the diagnosed electron densities of the EBIT-II
plasma from different line ratios. They are obtained by comparing
the theoretical line ratios with the experimental measurements
which are indicated by the filled diamonds in Fig. 1---3.
\begin{table*}
\centering
    \caption[1]{The diagnosed
    electron density of the EBIT plasma assuming the isothermal plasma
    with three different temperatures
     log$T_e$(K)=6.3, 6.5 and 6.7.}
    \vspace{0.1cm}
      \[
\begin{array}{l|llllll}
            \hline
            & \multicolumn{6}{c}{{\rm Electron \>density}\> n_e\>({\rm \times10^{10}
            cm^{-3}})}\\
  &  R_1 &  R_2 & R_3  & R_4  & R_5 & R_6 \\
            \hline
 {\rm log}T_e=6.3 & 5.1 & 6.0 & 7.6 & - & - & 45.5 \\
 {\rm log}T_e=6.5 & 5.6 & 9.5 & 9.0 & - & - & 55.7  \\
 {\rm log}T_e=6.7 & 6.1 & 14.4 & 10.7 & - & - & 95.8 \\
           \hline
         \end{array}
      \]
       \end{table*}
Given the uncertainties in the theoretical and measured line
ratios, the derived values of log~$n_e$ would be expected to be
accurate within $\pm0.3$~dex. The results corresponding to
log$T_e$(K)=6.5 shown in Table 2, indicate that the discrepancies
of diagnosed electron density from $R_1, R_2$ and $R_3$ do not
exceed $\pm0.25$~dex. The consistency of the diagnosed electron
density from $R_1, R_2$ and $R_3$ further confirms that our
results are reliable. $R_2$ predicts a relative higher electron
density compared to $R_1$ and $R_3$ at log$T_e$(K)=6.5. Part of
the reason may be due to the temperature sensitivity of $R_2$ as
shown in Fig. 6. A change in $T_e$ of 0.2\,dex leads to 10\%
variation in $R_2$ at $n_e=10^{11}$cm$^{-3}$, which can explain
the relative higher diagnosed electron density from it.
\begin{figure*}
\centering
\includegraphics[angle=-90,width=8cm,clip]{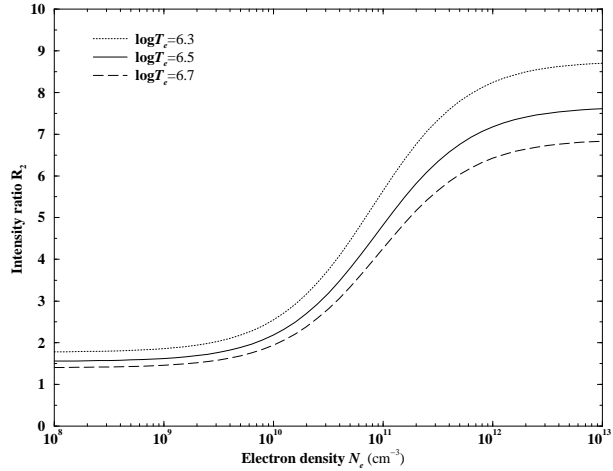}
\caption[short title]{The theoretical Ar XIV emission line ratio
$R_2$, plotted as a function of electron density ($n_e$ in
cm$^{-3}$) at temperature log$T_e$(K)=6.3 (dotted line), 6.5
(solid line) and 6.7 (dashed line)}
\end{figure*}
The measured value of $R_4$ (6.33) exceeds the high
density-sensitive limit, so no electron density can be derived as
shown by Fig. 2. The line at 31.360~\AA\, is strongly shadowed by
Ar XII line at 31.374~\AA\,~[22], therefore no reliable measured
line intensity is available, which results in large uncertainty in
$R_5$. The line at 32.014~\AA\, blends with weak lines of Ar XIII
at this wavelength, so a higher electron density is derived from
$R_6$.

Fig. 6 shows that the ratio $R_2$ is slightly sensitive to the
electron temperature. At a lower electron temperature
log$T_e$(K)=6.3, the derived electron density from $R_2$ can be
consistent with that from $R_1$ and $R_3$ within a much smaller
uncertainty range (25\%). Decreasing temperature from
log$T_e$(K)=6.5 to 6.3, the uncertainty range decreases from 78\%
to 25\% as shown in Table 2. Which indicates that the electron
temperature of the laboratory plasma is lower than that of maximum
Ar XIV fractional abundance in ionization equilibrium.

Table 3 lists the experimental and theoretical intensity ratios at
log$T_e$(K)=6.5 and  $n_e=5.6\times10^{10}$cm$^{-3}$. For ratios
$R_2$ and $R_6$, poor consistencies between the measurements and
theoretical calculations can be found in Table 3. The
temperature-sensitivity can explain the discrepancy for $R_2$,
while the contribution from those weakly Ar XIII lines around
32.014~\AA\, line can explain the deviation for $R_6$. The large
discrepancy for $R_4$ may be due to that the 30.344~\AA\, line is
strongly contaminated by Ar XIII lines located at this wavelength
region~[22].
\begin{table*}
\centering
    \caption[1]{Experimental and theoretical intensity ratios.
    The experimental values are from Lepson's work~[22]}
    \vspace{0.1cm}
      \[
\begin{array}{lll}
            \hline
{\rm Ratio}& {\rm Experiment} & {\rm Theory}  \\
            \hline
{\rm R_1}  &  0.95  & 0.96 \\
{\rm R_2}  &  4.75  & 3.87 \\
{\rm R_3}  &  4.75  & 4.81 \\
{\rm R_4}  &  6.33  & 2.35 \\
{\rm R_5}  &  -     & 6.13 \\
{\rm R_6}  &  3.17  & 2.10 \\
           \hline
         \end{array}
      \]
       \end{table*}

\section{Conclusion}
We found that intensity ratios $R_1$---$R_6$ of emission lines
arising from $n=3-2$ transitions of Ar XIV are sensitive to
electron density. And the ratio $R_1$ is insensitive to electron
temperature. For example, a change in $n_e$ of 1~dex leads to
variation with factor $\sim$4.4 in $R_1$, while a change in $T_e$
of 0.2~dex (i.e. 58\%) leads to a less than 1\% variation at
$n_e=10^9$ cm$^{-3}$, up to 3\% at $n_e=10^{11}$ cm$^{-3}$.
Therefore $R_1$ is very useful for the electron density
diagnostics of laboratory and astrophysical plasmas. The two lines
of Ar XIV relevant to $R_1$ are the strongest, and no blended Ar
lines have been detected in the experiment conducted by Lepson
\etal~[22] in EBIT-II. By comparison between the experimental
value and theoretical line intensity ratio $R_1$, an electron
density of $n_e=5.6\times10^{10}$cm$^{-3}$ of the laboratory
plasma is derived, which is consistent with value
$6.0\times10^{10}$cm$^{-3}$ recently diagnosed by Chen \etal~[34]
using triplet line of N VI. The derived electron densities from
$R_2$ and $R_3$, are also consistent with that within uncertainty,
which further confirms that our results are reliable. In addition,
the electron density derived from these ratios eliminates the
uncertainty from ionization balance and electron abundance
determination. At a lower logarithmic electron temperature
(log$T_e$(K)=6.3), the derived electron density from $R_2$ can be
consistent with that from $R_1$ and $R_3$ within a much smaller
uncertainty range (25\%). Which indicates the temperature of the
laboratory plasma is lower than log$T_e$(K)=6.5.

\section*{Acknowledgments}
This work was supported by the National Natural Science Foundation
of China under Grant No. 10373014 and 10403007, and the Chinese
Academy of Sciences under Grant No. KJCX2-W2.

\end{document}